\documentclass[letterpaper]{article}
\usepackage[english]{babel}
\usepackage[utf8]{inputenc}
\usepackage[colorinlistoftodos, color=green!40, prependcaption]{todonotes}
\usepackage{graphicx}
\usepackage{amsthm}
\usepackage{mathtools}
\usepackage{xcolor}
\usepackage[left=20mm,right=20mm,top=25mm,columnsep=15pt]{geometry} 
\usepackage{adjustbox}
\usepackage{placeins}
\usepackage[T1]{fontenc}
\usepackage{lipsum}
\usepackage{csquotes}
\usepackage{mathptmx}
\usepackage{etoolbox}
\usepackage{subcaption}
\usepackage{xparse}
\usepackage{xcolor}
\usepackage{siunitx}

\newcommand{\q}{\ensuremath{\bm q}}

\NewDocumentCommand{\gfun}{om}{\ensuremath{g_{#2\IfValueT{#1}{,#1}}(\q, \tau)}}
\NewDocumentCommand{\gone}{o}{\gfun[#1]{1}}
\NewDocumentCommand{\gtwo}{o}{\gfun[#1]{2}}
\newcommand{\ctwo}{\ensuremath{c_{2}(\q, t, \tau)}}
\newcommand{\I}[1][t]{\ensuremath{I(#1,p)}}
\newcommand{\avgb}[2]{\ensuremath{\left<#1\right>_{#2}}}

\usepackage[pdftex, pdftitle={Article}, pdfauthor={Author}]{hyperref} 
\usepackage{bm}
\usepackage{achemso}
\usepackage{authblk}

\begin{document}
\title{XPCS-Echo and broad relaxation measurements using a bunch-mode data acquisition scheme}
\author[1]{William Ch{\`e}vremont}
\affil[1]{ESRF, The European Synchrotron, F-38043 Grenoble, France}
\author[2]{Thomas Gibaud}
\affil[2]{ENSL, CNRS, Laboratoire de Physique, F-69342 Lyon, France}
\author[1]{Yuriy Chushkin}
\author[1]{Theyencheri Narayanan}
\date{\today, *E-mail: narayan@esrf.fr} 
\maketitle

\begin{abstract}
The analysis of echoes observed in the intensity-intensity autocorrelation functions [$g_2(q,\tau)$] is a powerful method for probing nonaffine deformations and yielding behavior of soft viscoelastic materials subjected to an oscillatory shear. Multispeckle X-ray photon correlation spectroscopy (XPCS) measurements of large number of echoes with high time resolution impose severe constraints in terms of the computational hardware and potential degradation of the sample. These issues can be alleviated by implementing a bunch-mode data acquisition scheme in which the highest resolution frames or bunches are centered at the echo peaks. In addition, by placing the bunches in an aperiodic Fibonacci sequence, $g_2(q,\tau)$ over a long time span can be measured with orders of magnitude lesser number of frames. The performance of these XPCS acquisition schemes is demonstrated using slowly relaxing model colloidal suspensions. Furthermore, an analytical expression is provided for the quantitative description of full $g_2(q,\tau)$.         
\end{abstract}

\section*{Keywords}
XPCS, XPCS-Echo, Rheo-XPCS, USAXS, multiscale dynamics, colloids,                          

\section{Introduction}
The Photon Correlation Spectroscopy (PCS) or Dynamic Light Scattering (DLS) has been widely used for investigating complex relaxation processes in a broad range of condensed matter systems \cite{Cummins1973,Brown1993,Berne2000}. The light scattering-Echo (LS-Echo) method is an extension of PCS to probe a sample subjected to an oscillatory motion or deformation \cite{Hebraud1997,Hohler1997}. For a perfectly elastic sample, periodic echoes can be observed in the intensity-intensity autocorrelation function [$g_2(q,\tau)$] that do not decay with time while the viscoelasticity is manifested by a gradual damping of echo peaks \cite{Petekidis2002,Pham2004}. The damping profile of the echo peaks contains information on the nonaffine contributions such as that due to Brownian diffusion. The LS-Echo method has been successfully used to detect nonlinear dynamics of aqueous foam \cite{Hohler1997}, yielding behavior of emulsion droplets \cite{Hebraud1997}, colloidal glasses \cite{Petekidis2002} and gels \cite{Laurati2014}, and even to obtain proper ensemble averaging \cite{Pham2004}. In all these studies, LS-Echo measurements were performed in the strongly multiple scattering regime using diffusing wave spectroscopy (DWS) \cite{Hohler1997,Hebraud1997}. As a result, the wave vector dependent information was not obtained. 

The X-ray analogue of PCS or X-ray photon correlation spectroscopy (XPCS) can in principle overcome the above shortcoming of LS-Echo \cite{Leheny2015}. However, the quality of the echo signal and time resolution were limited by the available photon flux and frame rate of the two-dimensional (2D) detector \cite{Rogers2014,Rogers2018}. With the advent of the fourth generation synchrotron light sources, that offer more coherent flux and availability of fast 2D X-ray detectors, the XPCS technique has become more attractive to study faster dynamics in soft matter systems \cite{Narayanan2024}. Main advantages of XPCS over DLS are the ability to investigate optically opaque samples, the use of a 2D detector enables multispeckle and direction-dependent analysis (e.g. along vertical and horizontal directions), and the access to a wider range of scattering vector ($\bf{q}$) simultaneously. The drawbacks however are X-ray induced dynamics and progressive degradation of the sample, and significantly lower scattering contrast for most soft matter samples. Therefore, DLS is more adapted for dilute samples, while XPCS becomes more suitable for dense and opaque samples \cite{Kamal2024}.

Nowadays, the PCS is more widely used as a routine characterization tool that relies on the diffusive dynamics within the sample.  But the PCS can also be employed for velocimetry \cite{Pike1977}, characterization of shear flows \cite{Fuller1980} and obtain the small-scale statistical properties of turbulent flows \cite{Tong1988,Pak1991}. Similarly XPCS can also be utilized to probe shear flows at moderate rates \cite{Leheny2015}, with the ability to decipher the parallel and perpendicular components \cite{Westermeier2016,Chevremont2025}. This particular aspect of multispeckle XPCS combined with the oscillatory shear deformation is exploited in the echo method presented below. In conventional XPCS measurements, the speckle patterns are recorded with equal spacing in time, which limit both the shortest and longest lag time ($\tau$) sampled in a single acquisition within the available hardware resources. In PCS this limitation has overcome by means of a nonlinear sampling scheme involving multi-$\tau$ correlator and quasi-logarithmic time grid \cite{Schatzel1987}. More recently, multi-$\tau$ two-time correlation \cite{Brugnara2026} and logarithmic sampling \cite{Hoshino2026} schemes have been implemented for XPCS. The bunch-mode correlation method presented here is clearly distinct from those approaches.     

\section{Materials and method}
This Section describes the experimental details, measurement procedure and theoretical background for data analysis.  
\subsection{Colloidal suspensions}
Two different colloidal systems consisted of either  sterically-stabilized polymethyl methacrylate (PMMA) latex particles with mean radius, $R_S \simeq$ \SI{380}{nm} and size dispersity $\simeq 5\%$ suspended in cis-decalin \cite{Antl1986} or stearyl silica particles $R_S \simeq$ \SI{67}{nm} and size dispersity $\simeq 9\%$ suspended in n-dodecane \cite{Narayanan2006}. The respective volume fractions ($\phi$) were \SI{0.58}{} and \SI{0.18}{}, corresponding to the former in a repulsive colloidal glassy \cite{VanMegen1991} and the latter in an attractive gel \cite{Sztucki2006} states at ambient temperature. These $\phi$ values were estimated from the scattered intensity, $I(q)$, on an absolute scale \cite{DiCola2009}. A larger particle size was chosen for the echo measurements to obtain good scattering signal within a short acquisition time. The gel sample was contained in a sealed flat glass capillary of thickness \SI{0.5}{mm}. The sample temperature was varied using a Mettler-Toledo heating stage (HS82).  
\subsection{Rheometer setup}
The oscillatory shear was applied using a Haake RheoStress 6000 rheometer equipped with a capillary coaxial shear cell \cite{Narayanan2020}. The shearing geometry consisted of thin-walled (\SI{0.05}{mm}) quartz capillaries of inner diameter \SI{2}{mm} as the stator and outer diameter \SI{1}{mm} as the rotor. This provided a shear gap of \SI{0.5}{mm} and the height of the cell was about \SI{35}{mm}. The XPCS data acquisition was hardware synchronized with the onset of oscillatory motion of the rheometer rotor.
\subsection{X-ray scattering}
The experiment setup and data acquisition hardware were the same as in a previous publication \cite{Chevremont2025}. Reported developments were carried out at the TRUSAXS instrument of the ESRF (beamline ID02) \cite{Narayanan2022}. Multispeckle XPCS measurements were conducted over the ultrasmall angle (UA) region with an incident X-ray energy of $\SI{12.230}{keV}$ (wavelength, $\lambda \simeq $ \SI{1.013}{\angstrom}) and a sample-to-detector distance of \SI{30.7}{m}. The setup used a 2D hybrid photon-counting pixel array detector, EIGER2 500K (PSI), with a maximum frame rate of \SI{23}{kHz} \cite{Chevremont2024}. Measurements covered a $q$ range, \SI{1.7e-3}{\per\nano\meter} $\leq q \leq$ \SI{1.e-1}{\per\nano\meter}, where $q$ is the magnitude of the scattering vector, given by $q=(4 \pi / \lambda) \sin(\theta/2)$, with $\theta$ the scattering angle. 
\subsection{Bunch generation}
\begin{figure}[h]
\centering
(a)\includegraphics[width=0.6\linewidth]{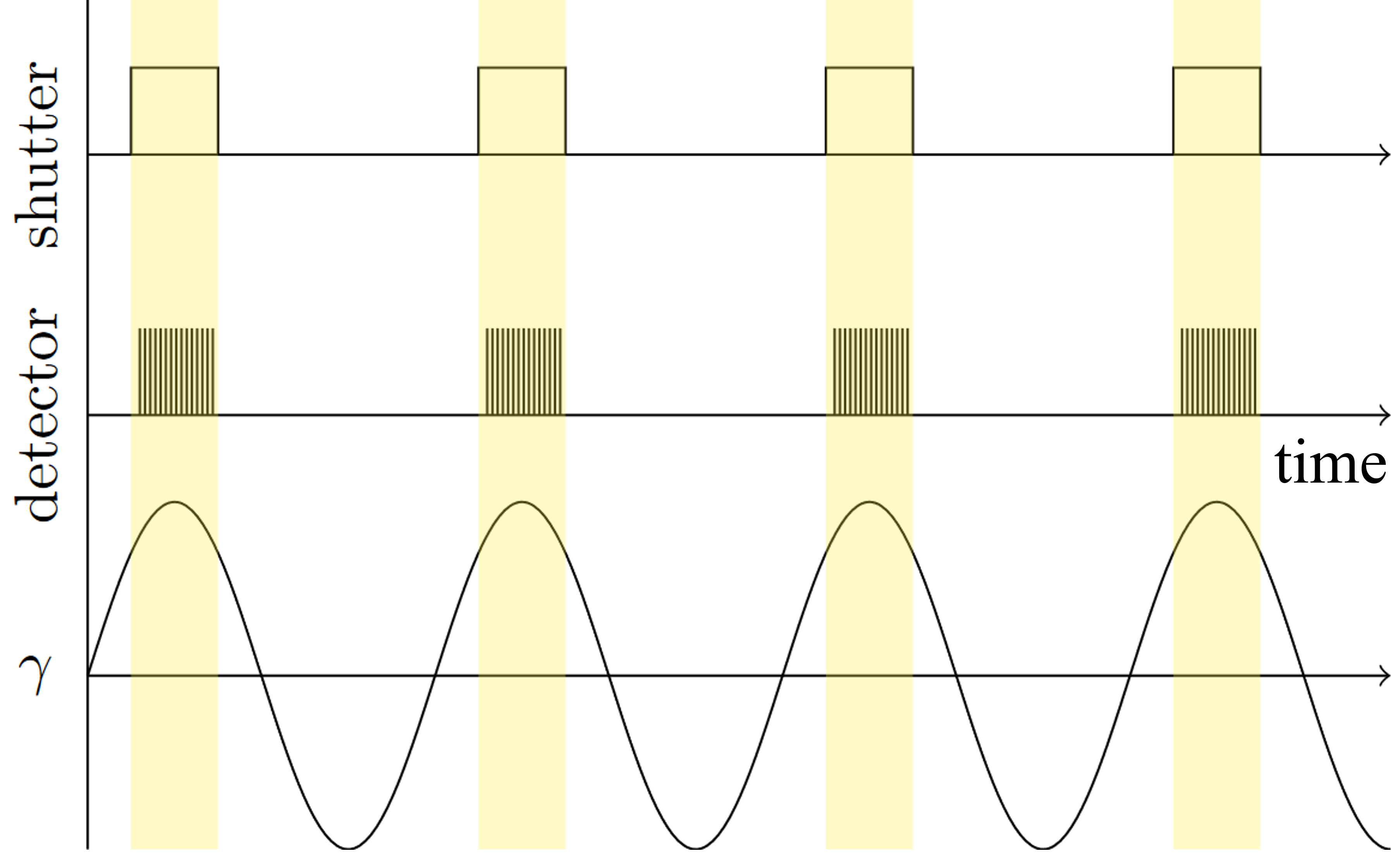}
(b)\includegraphics[width=0.6\linewidth]{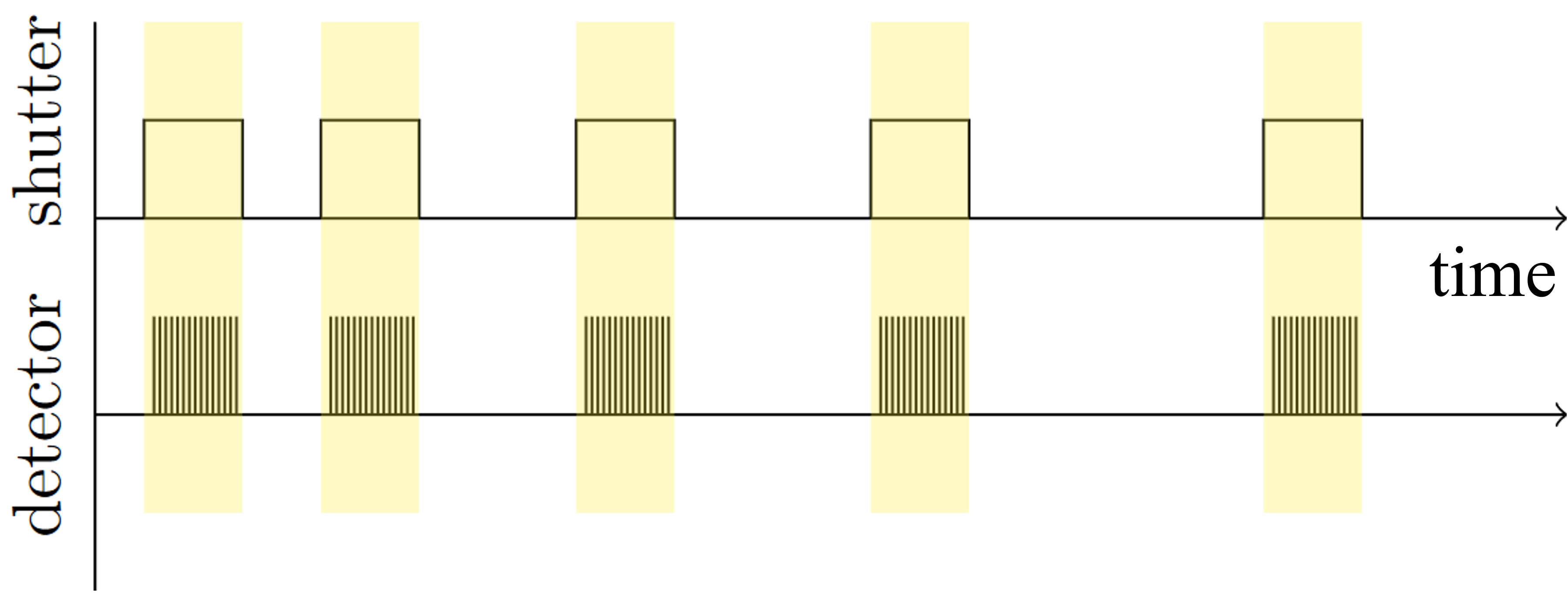}
	\caption{Schematic diagrams of the pulse trains used in (a) echo and (b) bunch-mode aperiodic sequence XPCS measurements. In the former, the pulse trains are synchronized with the oscillatory deformation ($\gamma$) and in the latter the sequence needs to ensure sufficient overlap in lag time between successive bunches.}
\label{fig:BunchScheme}
\end{figure}

The bunch-mode here refers to a train of bunched TTL pulses used for triggering the detector acquisition, and not directly related to the electron bunches in the storage ring of the synchrotron. The desired TTL pulse trains were generated using the ESRF developed digital electronic device Multipurpose Advanced Electronics for Sequencing, Triggering and Reconfigurable I/O (MAESTRIO) \cite{Maestrio}. This instrument enabled an easy programming of the bunch patterns via the beamline control software (BLISS). The output TTL pulses of the MAESTRIO were used to gate the X-ray detector, controller of the fast beam shutter that remains open only during a bunch and other ancillary devices. The sequence is initiated when the rheometer sends a trigger signal to the MAESTRIO external trigger input at the onset of shear deformation ($\gamma$). By matching the oscillation frequency ($f$) and an initial delay, the bunches were centered around the maximum of $\gamma$. Figure \ref{fig:BunchScheme} schematically depicts the pulse trains used for XPCS-Echo (a) and quasi-logarithmic or aperiodic (b) data acquisition for covering a broad range in lag time ($\tau$). Within a bunch the detector acquires at the highest frame rate. In the aperiodic acquisition scheme, the gap between two successive bunches should be significantly smaller than the cumulative $\tau$ of previous bunches in order to ensure sufficient temporal overlap. A Fibonacci sequence was found to fulfill this requirement.          
\subsection{XPCS analysis}
In a multispeckle XPCS measurement, the temporal fluctuations of scattered intensity provide access to the time-dependent dynamics within the sample via the (normalized) two-time intensity-intensity autocorrelation function [\ctwo, TTCF] \cite{Lehmkuhler2021,Chevremont2024}. 
\begin{gather}
    \ctwo = \dfrac{\avgb{\I[t_1]\I[t_1+\tau]{}}{p}}{\avgb{\I[t_1]}{p}^2}, \forall p \in \q 
\end{gather}
Where $\I[t_1]$ is the intensity recorded by pixel $p$ at time $t_1$, $\tau$ is the lag time as mentioned before, and $t$ is the age defined as the average time of the two frames ($t = (t_1 + t_2)/2 = t_1 + \tau/2$), $t_2=t_1 + \tau$ being the time of the second frame. \avgb{\cdot}{p} and \avgb{\cdot}{t} are the averaging operations over pixels and time, respectively. The  pixels correlated are within a region of interest (ROI), characterized by its average $q$ value, and restricted in azimuthal angle when there is an anisotropy as in shear deformation.
The conventional intensity-intensity autocorrelation function [\gtwo] is the time-average of this TTCF.
\begin{gather}
    \gtwo = \avgb{\ctwo}{t}
\end{gather}
Here, \gtwo{} is related to the field-field autocorrelation function [\gone] via the Siegert relation \cite{Berne2000,Brown1993}:
\begin{gather}
   \gtwo = 1 + g_0 \left|\gone\right|^2
\end{gather}
where $g_0$ is the speckle contrast, which is determined by the coherence properties of the X-ray beam and angular resolution of the instrument  \cite{Lehmkuhler2021,Chevremont2024}.

To model \gone of a suspension subjected to shear deformation, several contributions need to be taken into account: the Brownian diffusion of particles, the relative Doppler shifts caused by the shear deformation and the decorrelation effect due to particles transiting the scattering volume by the flow. These effects can be factorized in the combined field-field autocorrelation function as given below \cite{Busch2008,Burghardt2012}: 
\begin{gather}
    |\gone|^2 = |\gone[D]|^2|\gone[F]|^2|\gone[T]|^2 \label{eq:g1_comb}
\end{gather}
The first term \gone[D] accounts for the Brownian motion, which exhibits an exponential or a similar decay \cite{Berne2000}:
\begin{gather}
    |\gone[D]|^2 = \exp\left(-2D q^2 \tau\right)\label{eq:g1_D}
\end{gather}
where $D$ is the diffusion coefficient given by the Stokes-Einstein relation, 
\begin{gather}
D = \dfrac{k_BT}{6\pi R_H\eta_f} \label{eq:stokes-einstein}
\end{gather}
where $k_B$ is the Boltzmann constant, $T$ is the absolute temperature, $R_H$ is the hydrodynamic radius of particles and $\eta_f$ is the dynamic viscosity of the solvent at $T$.

The second term \gone[F] describes the flow field followed by the particles, more specifically the average Doppler shift introduced by the flow field along the beam direction x \cite{Berne2000,Chang1981}. In an oscillatory shear, this Doppler shift is time ($t$) dependent and therefore  
\begin{gather}
g_{1,F}(\bm q, t, \tau) = \left< \exp(i\bm q\cdot\bm{\Delta r}) \right>_x
\end{gather}
The velocity field for an oscillatory motion in a Couette geometry, 
\begin{gather}
    \bm v(t) = \gamma \omega x \sin(\omega t)\bm e_y
\end{gather}
with $\bm e_x$, $\bm e_y$ and $\bm e_z$ being unit vectors along the beam, horizontal and vertical directions, respectively and $\omega=2 \pi f$. The instantaneous position is given by $\bm r(t) = \int_0^t \bm v(t') dt'$.
\begin{gather}
    \bm{\Delta r}(t,\tau) = 2 \gamma x\sin\left(\omega t + \dfrac{\omega\tau}{2}\right)\sin\left(\dfrac{\omega\tau}{2}\right)\bm e_y
\end{gather}
The accumulated phase shift is given by
\begin{gather}
    \bm q\cdot\bm{\Delta r} = 2 q\cos(\psi)\gamma x\sin\left(\omega t + \dfrac{\omega\tau}{2}\right)\sin\left(\dfrac{\omega\tau}{2}\right)
\end{gather}
where $\psi$ is the azimuthal angle. For a uniform distribution of particles over the Couette gap, $L$ , 
\begin{gather}
    g_{1,F}(\bm q, t, \tau) = 
    \dfrac{1}{L}\int_{0}^{L}\exp\left(2iq\cos(\psi)\gamma x\sin\left(\omega t + \dfrac{\omega\tau}{2}\right) \sin\left(\dfrac{\omega\tau}{2}\right) \right)dx\\ \nonumber
         = \text{sinc}\left[q\gamma L \cos(\psi)\sin\left(\omega t + \dfrac{\omega\tau}{2}\right)\sin\left(\dfrac{\omega\tau}{2}\right)\right] 
     \exp\left[iq\gamma L\cos(\psi)\sin\left(\omega t + \dfrac{\omega\tau}{2}\right)\sin\left(\dfrac{\omega\tau}{2}\right)\right]
\end{gather}
Since the normalization of the term in complex exponential becomes $1$, 
\begin{gather}
  |g_{1,F}(q, t, \tau)|^2 = \text{sinc}^2\left[\dfrac{q}{\pi}\cos(\psi)\gamma L\sin\left(\omega t + \dfrac{\omega\tau}{2}\right)\sin\left(\dfrac{\omega\tau}{2}\right)\right]
\end{gather}
For low shear rates and the beam size used here ($\simeq$\SI{25}{\mu m}), $|\gone[T]|^2 \simeq$ 1. However, it becomes important for higher $\gamma$ and $\omega$ as given by,
\begin{gather}
|\gone[T]|=\int_0^L \exp\left(-\dfrac{\gamma^2 x^2 \sin^2\left( \dfrac{\omega\tau}{2}\right)}{\sigma_y^2}\right) dx
\end{gather}
where $\sigma_y$ is the Gaussian beam size. The measured $g_2(q, \tau)$ is the average of $g_2(\bm q, t, \tau)$ [$\equiv \ctwo$] over the total measurement time, $t_{exp}$, 
\begin{gather}
g_2(q, \tau) = \dfrac{1}{t_{exp}} \int_0^{t_{exp}}g_2(\bm q, t, \tau)dt
\end{gather}
For sufficiently long acquisition duration, $i.e$. when the weight of a partial oscillation does not change the average significantly, the integral becomes independent of the acquisition time. Then for the bunch-mode acquisition:
\begin{gather}
g_2(q, \tau) = \dfrac{1}{N_b}\sum_{k=0}^{N_b}\dfrac{1}{t_b} \int_{kt_{d}}^{kt_d+t_b}g_2(\bm q, t, \tau)dt
\end{gather}
with $N_b$ the number of bunches, $t_b$ the bunch duration, and $t_d$ the delay between bunches.

\section{Results and discussion}
This section illustrates the main features of bunch-mode acquisition and discusses the key advantages and limitations.
\subsection{XPCS-Echo measurements}
\begin{figure}[h]
    \centering
    \includegraphics[width=0.8\textwidth]{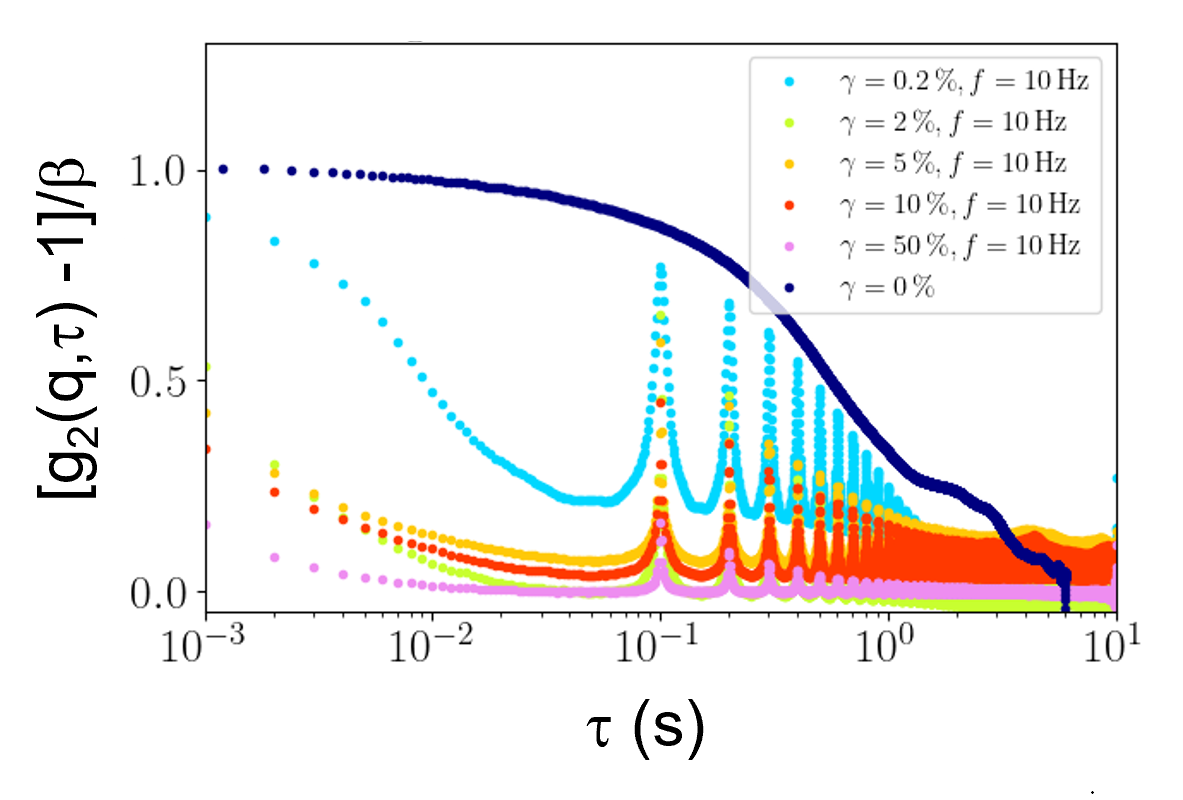}
	 \caption{Typical echoes observed in the autocorrelation function of a dense PMMA colloidal suspension in the glassy state ($\phi \simeq$ \SI{0.58}{}) as a function of oscillatory strain with $f$=\SI{10}{Hz} at $q \simeq$ \SI{4e-3}{nm^{-1}}. The faster damping of the echoes with $\gamma$ signifies the yielding of the sample above a critical deformation.}
    \label{fig:echo}
\end{figure}

In an XPCS-Echo measurement, the sample is subjected to a periodic oscillatory motion such as in an oscillatory deformation ($\gamma$) at a fixed frequency ($f$). The echoes in $g_2(q,\tau)$ appear when the sample returns to exactly the initial position. Figure \ref{fig:echo} depicts the evolution 
of  $g_2(\tau, q)$ for a viscoelastic sample as a function of the amplitude of oscillatory shear deformation with $f$=\SI{10}{Hz} along the flow direction. The sample consisted of a dense suspension of PMMA particles in cis-decaline ($\phi \simeq$ \SI{0.58}{}) in the vicinity of its colloidal glass transition. At this frequency and relaxation rates, the conventional XPCS acquisition with linearly spaced frames (10000) captures the decay of $g_2(q, \tau)$ at rest and upon deformation. The initial decay of $g_2(q, \tau)$ with increasing $\gamma$ is due to the Doppler shift caused by the flow. The faster damping of the echoes with increasing $\gamma$ signifies the nonaffin deformations induced by the oscillatory shear and yielding of the sample.  

\begin{figure}[h]
    \centering
    \includegraphics[width=0.8\textwidth]{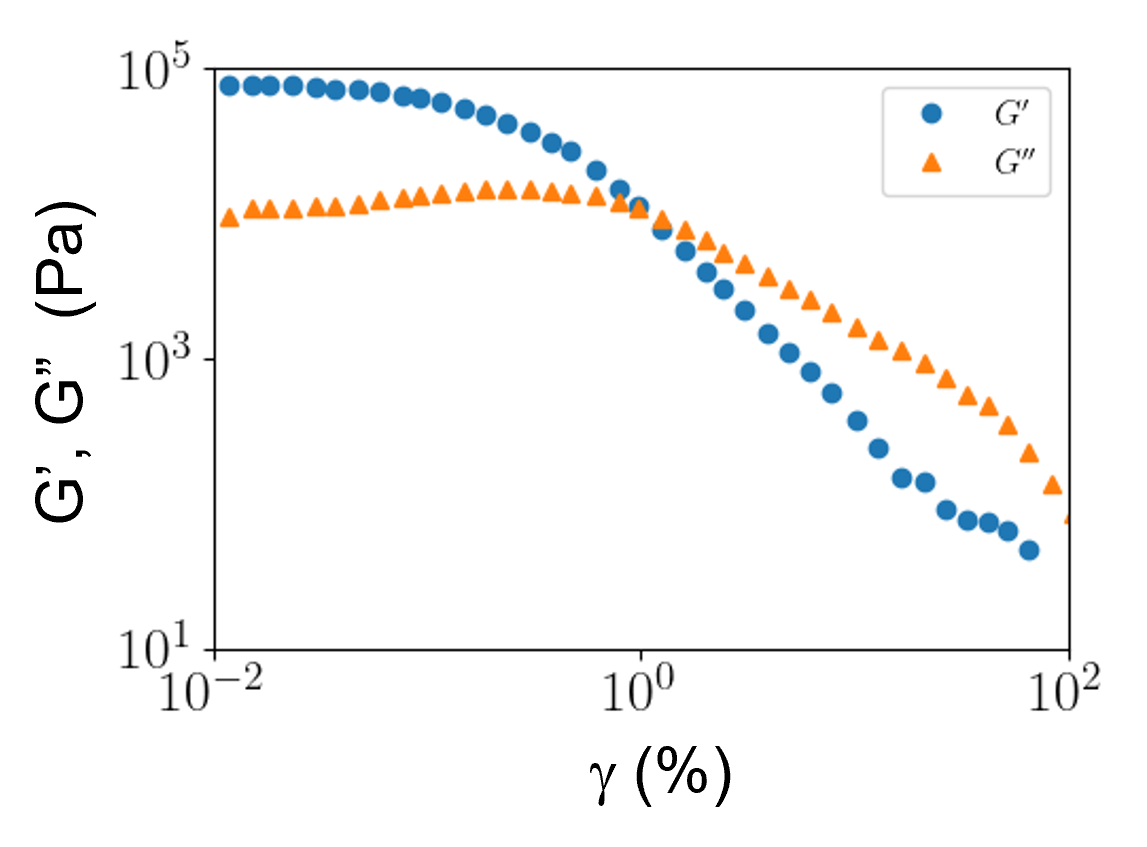}
    \caption{Typical viscoelastic behavior of the dense PMMA colloidal suspension ($\phi \simeq$ \SI{0.58}{}) in the glassy state measured using a standard plate-plate geometry. The sample yielded above $\gamma \sim$\SI{1}{\%}.}
    \label{fig:rheology}
\end{figure}

\begin{figure}
    \centering
    \includegraphics[width=0.7\linewidth]{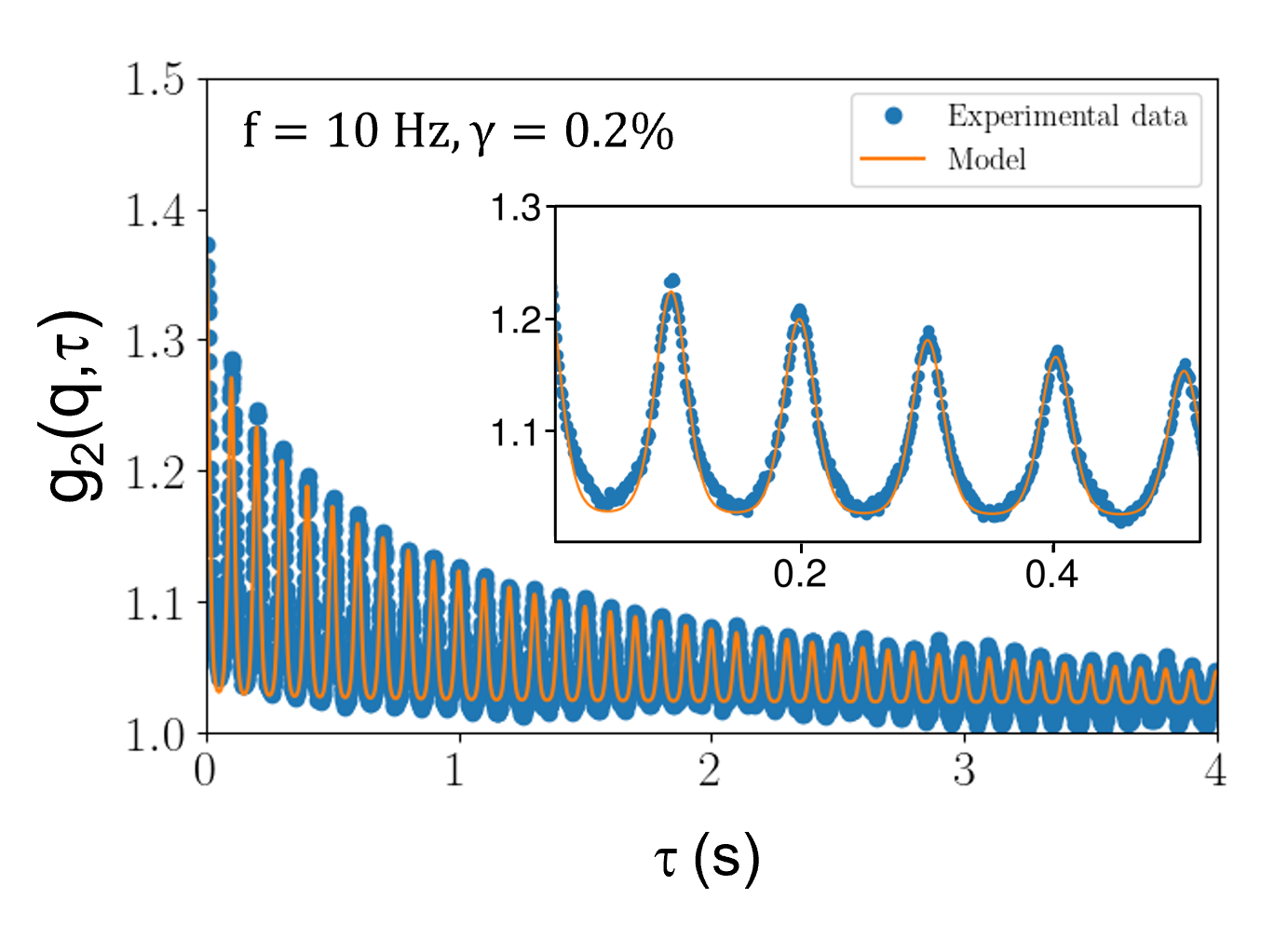}
	\caption{The modeling of echoes for $\gamma$=\SI{0.2}{\%} and $f$=\SI{10}{Hz} using Equation (16) for the PMMA suspension at $q \simeq$ \SI{4e-3}{nm^{-1}}. The inset depicts the higher resolution description of the first five peaks.  $g_2(q,\tau)$ does not fully decay to 1 indicating non-zero value of $f(q,\infty$).}
    \label{fig:echo_fit}
\end{figure}

Figure \ref{fig:rheology} presents the rheological behavior (elastic and viscous shear moduli, $G'$ and $G"$, respectively as a function of $\gamma$) of the sample measured using a standard plate-plate geometry. The yield point is located around $\gamma \simeq$ \SI{1}{\%}. This is consistent with the faster damping of echoes in Figure \ref{fig:echo}.   
It can be noticed that $g_2(\tau, q)$ in Figure \ref{fig:echo} did not decay to $1$ even well-above the yield point ($\gamma \simeq$ \SI{50}{\%}). This non-decaying part has been attributed to frozen-in or dynamically arrested component in a glassy system \cite{VanMegen1991}. This frozen-in component is not subjected to nonaffin deformation and therefore manifests as undamped echoes. This constant echo signal can be accounted by adding a nonergodicity parameter, $f(q,\infty)$, to $g_1(q,\tau)$ in Equation (\ref{eq:g1_D}). $f(q,\infty)$ represents the frozen-in component that does not contribute to the decay of the intermediate scattering function \cite{VanMegen1991}. The resulting expression has the following form:
\begin{gather}
  g_{2}(q, \tau) = 1 + \dfrac{g_0}{t_{exp}} 
  \int_0^L \exp\left(-2\dfrac{\gamma^2 x^2 \sin^2\left( \dfrac{\omega\tau}{2}\right)}{\sigma_y^2}\right) dx
\int_0^{t_{exp}}\text{sinc}\left[\dfrac{q}{\pi}\cos(\psi)\gamma L\sin\left(\omega t + \dfrac{\omega\tau}{2}\right)  \sin\left(\dfrac{\omega\tau}{2}\right)\right]^2 dt 
\\ \nonumber \times 
\left[\exp \left (-\left(\dfrac{\tau}{\tau_a}\right)^\beta\right) + f(q,\infty) \right]^2 
\label{eq:ACF}
\end{gather}
Where $\tau_a$ and $\beta$ are the apparent relaxation time and stretching exponent describing the diffusive or nonaffin deformation, respectively. The mean relaxation time, $\tau_c = \langle\tau_a \rangle = (\tau_a/\beta)\Gamma(1/\beta)$, with $\Gamma$ the gamma function. $\tau_c$  is related to $D$ as $1/\tau_c \simeq D q^2$. Figure \ref{fig:echo_fit} depicts the modeling of $g_{2}(q, \tau)$ by means of Equation (16) with $\tau_a$, $\beta$ and $f(q,\infty)$ as fit parameters. The beam crossing term was found to be negligible at that $\gamma$. The description is satisfactory up to the full decay of $g_{2}(\tau, q)$ around $\tau \simeq$ \SI{4}{s}.    

\begin{figure}
    \centering
    \includegraphics[width=0.8\linewidth]{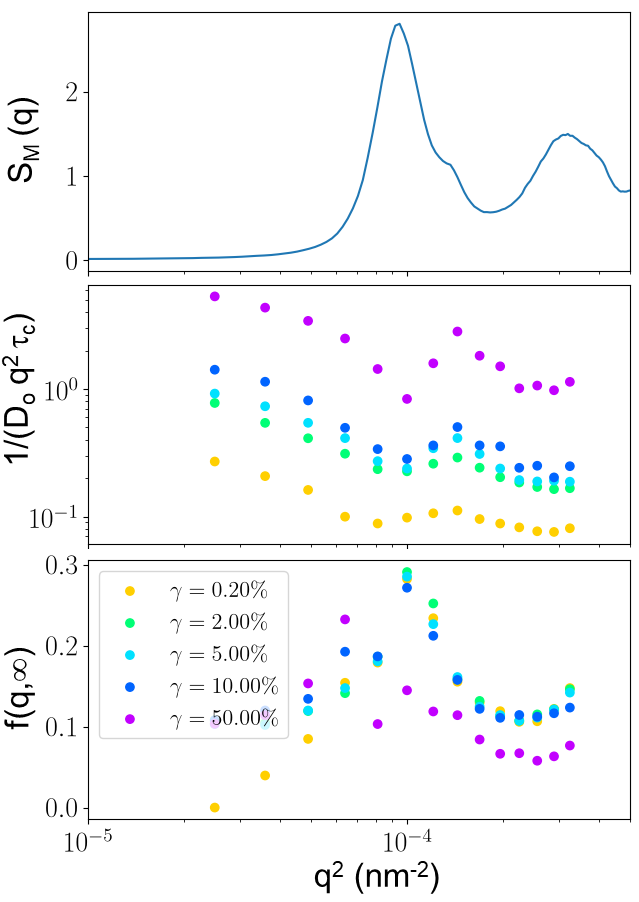}
	\caption{Dimensionless average relaxation rate and nonergodicity parameter as a function of $q^2$ for different $\gamma$ from the analysis using Equation (16) for the PMMA suspension. For a comparison the experimental effective structure factor, $S_M(q)$, is also shown. Notice the slowing down of the relaxation rate and increase of $f(q,\infty)$ around the peak of $S_M(q)$.}
    \label{fig:echo_fit_params}
\end{figure}

Figure \ref{fig:echo_fit_params} presents the main results of the analysis for selected $\gamma$ values. The exponent $\beta$ varied from \SI{0.88}{} at $\gamma$ = \SI{0.2}{\%} to \SI{0.33}{} at $\gamma$ = \SI{50}{\%}. To remove the regular $q$-dependence of the mean relaxation rate ($1/\tau_c$), it has been normalized by the non-interacting Brownian relaxation rate, $D_0 q^2$. 
For a comparison the experimental effective structure factor, $S_M(q)$, obtained from the division of normalized $I(q)$ by the form factor measured with a dilute sample, is also shown. As expected, the relaxation rate manifests a slowing down at the peak of $S_M(q)$. This implies that the dominant structural relaxation becomes sluggish at the equilibrium interparticle separation  \cite{Rogers2014}. On the other hand, $f(q,\infty)$ displays a maximum around the $S_M(q)$ peak as observed in light scattering studies of this colloidal glass \cite{VanMegen1991}, which is likely a consequence of the slower structural relaxation. The feature around $q \sim$\SI{0.019}{nm^{-1}} in $S_M(q)$ is a size dispersity effect at the first minimum of the spherical form factor \cite{DiCola2009}.  Overall, the relaxation rate increased and $f(q,\infty)$ varied non-monotonically with $\gamma$. In general, shear flow enhances the relaxation rate when the P{\'e}clet number is sufficiently large and the presented trend is consistent with an earlier work on PMMA colloidal glass using LS-Echo \cite{Petekidis2002}. The lack of a systematic variation of $f(q,\infty)$ with $\gamma$ further indicates that the frozen-in component is not subjected to nonaffin deformation.      

\begin{figure}
    \centering
    \includegraphics[width=0.8\linewidth]{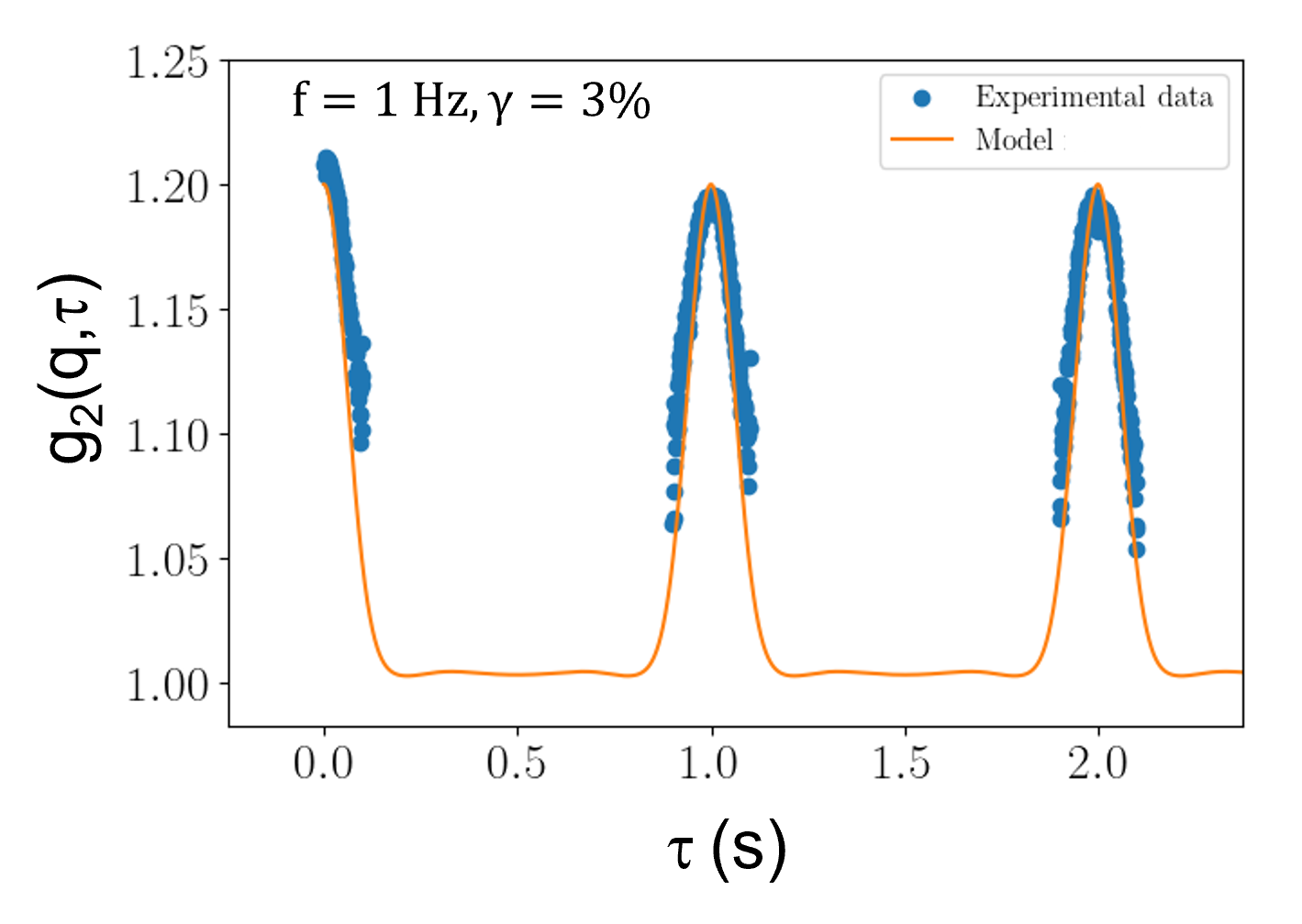}
    \caption{Echoes measured in bunch-mode using an aluminum rotor at a lower frequency \SI{1}{Hz} and their modeling in terms of Equation (16). In this case, the sample is perfectly elastic and the echoes do not damp with time.}
    \label{fig:echo_bunch_fit}
\end{figure}

The results presented above demonstrate that XPCS-Echo measurements using a linear acquisition sequence enabled quantitative extraction of the dynamical parameters of a dense colloidal suspension subjected to an oscillatory deformation of relatively high frequency. This approach becomes insufficient for slower relaxations involving weaker deformations. In that case the sample yields much more slowly and very long measurements are required to cross the yield point. Figure \ref{fig:echo_bunch_fit} shows the case for undamped echoes measured using an aluminum rotor of the rheometer instead of the capillary cell with $f=$\SI{1}{Hz} and $\gamma=$\SI{3}{\%}. In this case, the specimen is perfectly elastic, and both the diffusive and beam-crossing terms can be safely neglected. Using the bunch-mode acquisition with frames placed around the echo peaks, the required number of frames is significantly reduced while maintaining high time resolution of the measurement. Typically, 100 frames/echo at the highest frame rate enabled the recording of a large number of echoes with an order of magnitude lesser data. The peaks are well-described by Equation (16) using the known frequency of the oscillation. Further application of bunch-mode XPCS-Echo method is presented in the case of carbon black colloidal gels in Ref \cite{Chevremont2026}.        

Although the radiation damage was not a significant issue for the large PMMA particles due to their relatively high scattering power, the bunch-mode acquisition becomes useful even for rapidly damping echoes if the sample is susceptible to accumulated radiation dose. E.g. the data shown in Figure \ref{fig:echo} involving 100 echoes, can be measured with only 2000 frames instead of 10000 and thereby reducing the radiation dose by a factor \SI{5}{}. This factor inversely increases as the frequency of oscillation decreases. This aspect is illustrated in the next Section using a slowly relaxing colloidal gel sample. 

\subsection{Bunch-mode aperiodic acquisition}
The advantages of aperiodically spaced bunch-mode acquisition is illustrated by means of a colloidal suspension consisting of stearyl silica particles suspended in n-dodecane. This system behaves as hard-sphere repulsive colloids above a certain temperature (\SI{40}{^o C} in this case) and short-range attractive upon cooling by a fraction of a \SI{}{^o C} \cite{Narayanan2006}. A slowly relaxing and non-aging system can be realized by cooling the sample a few \SI{}{^o C} below the repulsive to short-range attractive transition temperature. 

\begin{figure}
    \centering
    \includegraphics[width=0.8\linewidth]{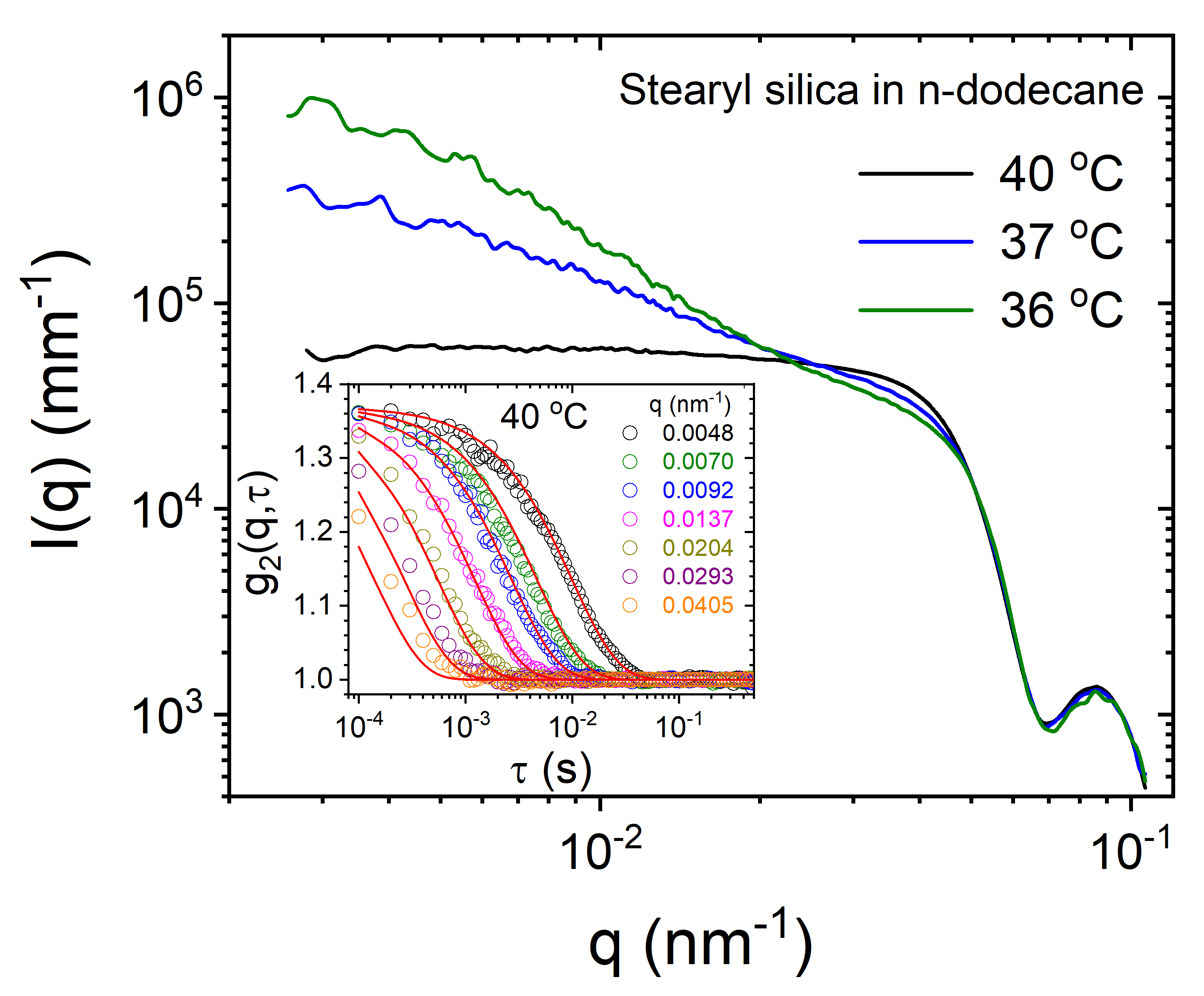}
	\caption{Normalized USAXS profiles from the stearyl silica suspension ($\phi \simeq$ \SI{0.18}{}) in the hard-sphere repulsive and short-range attractive gel states. The upturn in $I(q)$ at low $q$ indicates the formation of micron sized clusters below the transition temperature. The inset depicts the $g_2(q,\tau)$ traces in the hard-sphere region indicating the diffusive dynamics with $D \simeq$\SI{2.2}{\mu m^2 .s}.}
    \label{fig:USAXS}
\end{figure}

Figure \ref{fig:USAXS} displays the normalized ultrasmall-angle X-ray scattering (USAXS) profiles in the hard-sphere repulsive state at \SI{40}{^o C} and when the particles became attractive and formed clusters for a sample of $\phi \simeq$\SI{0.18}{}. The wiggles in the lowest $q$ region are due to insufficient time-averaging of the slowly fluctuating speckles within an acquisition time of \SI{0.1}{s}. The inset depicts diffusive dynamics of the particles described by Equation (\ref{eq:g1_D}) with $D \simeq$\SI{2.2}{\mu m^2.s} at \SI{40}{^o C}. These XPCS measurements were performed using linearly spaced frames at \SI{10}{kHz} rate. 

\begin{figure}
    \centering
    \includegraphics[width=0.8\linewidth]{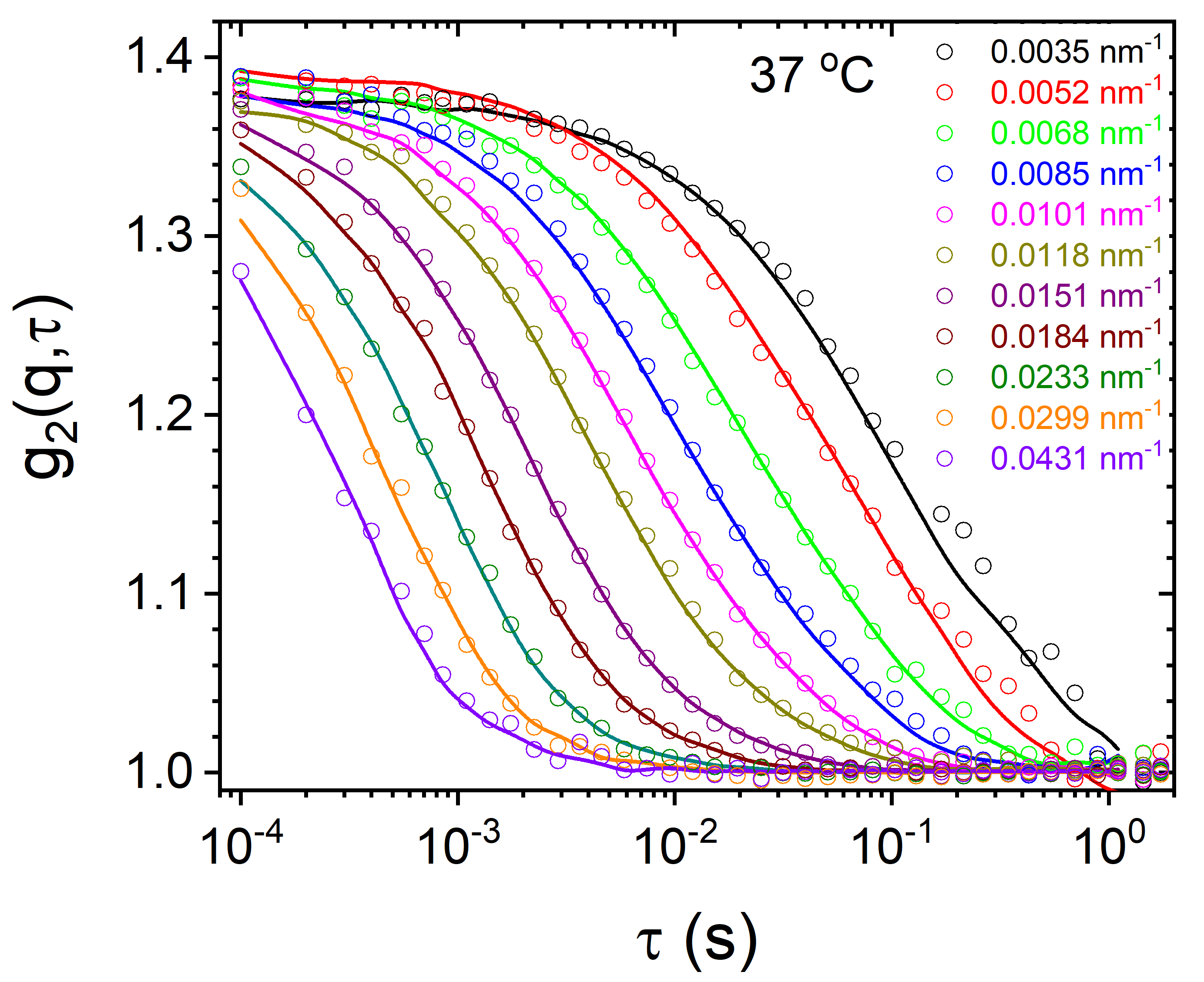}
	\caption{Comparison of $g_2(q,\tau)$ functions from the stearyl silica suspension ($\phi \simeq$ \SI{0.18}{}) in the clustered state measured using the conventional linearly spaced frames (continuous lines) and by means of the  bunch-mode (symbols). The X-ray dose for the latter was five times lower. The good agreement between two measurements establishes the equivalence of both schemes.}
    \label{fig:Comparison}
\end{figure}

As the sample was cooled gradually by \SI{3}{^oC}, $I(q)$ profile shows a strong upturn signifying the formation of large colloidal clusters. Figure \ref{fig:Comparison} presents the corresponding dynamics measured by linearly spaced frames (12000 at 10 kHz rate) and using the bunch-mode (12 bunches of 200 frames at 10 kHz rate). That is a factor of five lesser frames in the bunch-mode where the bunches were spaced aperiodically following a Fibonacci series. Therefore, the delay between two successive bunches was the sum of delays of two previous bunches, $t_d[n] = t_d[n-2]+t_d[n-1]$. The good agreement between data from the two measurements confirms the equivalence of bunch-mode acquisition for probing the dynamics in this sample. Even 100 frames per bunch did not make a significant difference. Due to contributions from both clusters and particles within them, $g_2(q,\tau)$ are no longer single exponential decay functions unlike in the repulsive state.         

\begin{figure}
    \centering
    \includegraphics[width=0.8\linewidth]{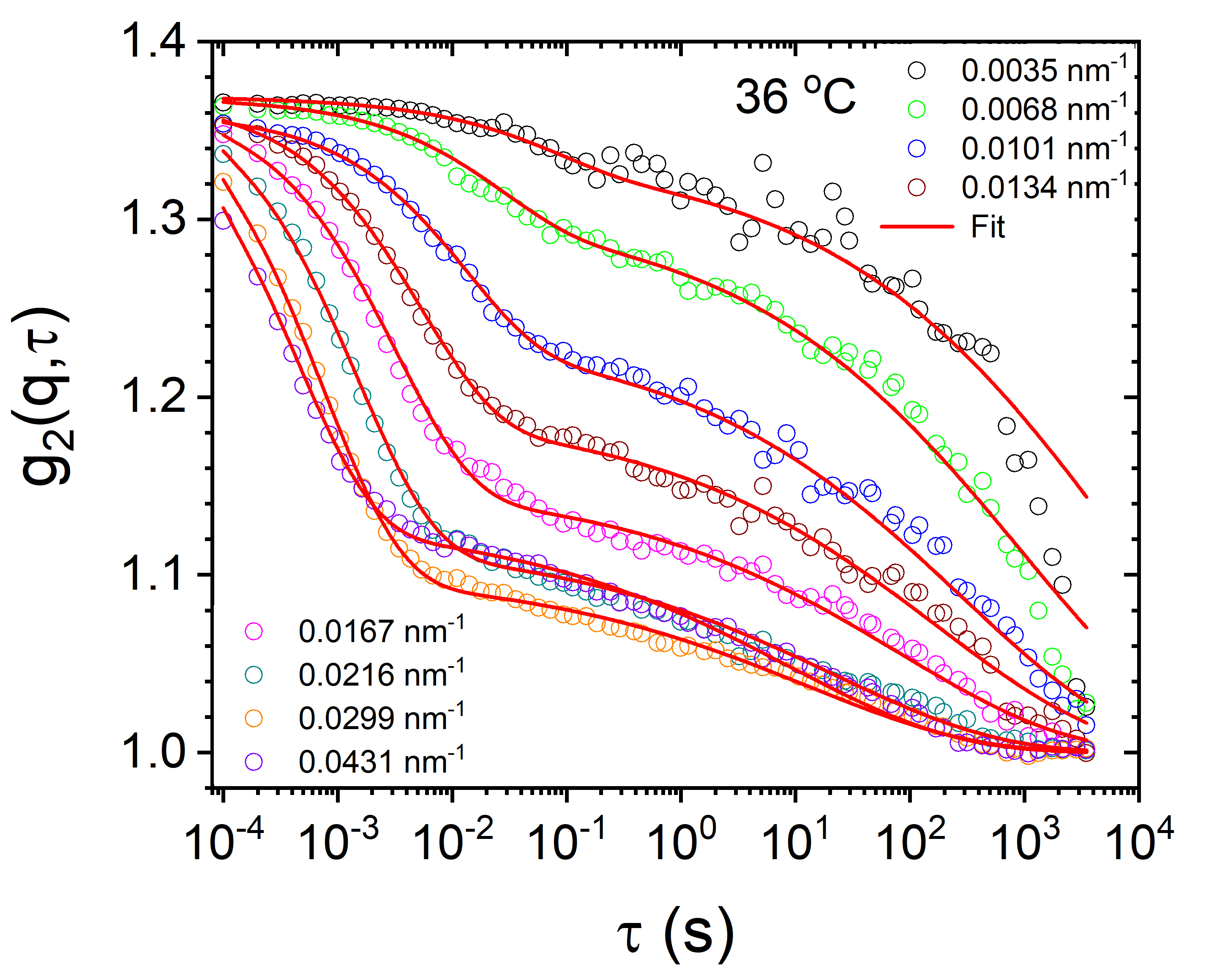}
    \caption{Very slowly relaxing $g_2(q,\tau)$ from the stearyl silica suspension in the deeply clustered state measured using the  bunch-mode acquisition (symbols). The continuous lines are modeling in terms of the two-step relaxation in Equation (\ref{eq:double_exponential}). Notice the colossally large time range probed in a single acquisition.}
    \label{fig:SlowRelax}
\end{figure}

Upon further cooling the sample by \SI{1}{^o C}, the relaxation slowed down dramatically as depicted in Figure \ref{fig:SlowRelax}. In this case, the conventional linear acquisition has become nearly impractical. The measurement shown involved 28 bunches in a Fibonacci series with 100 frames/bunch at 10 kHz rate. The fast decay at the initial stage cannot be captured without the highest rate acquisition. The two-step relaxation is attributed to the constrained motion of the particles within the cluster and the percolation of the clusters themselves \cite{Verduin1995}. The $g_2(q,\tau)$ functions can be described by the following phenomenological expression involving two stretched exponential terms.
\begin{equation}
g_2(q,\tau)=1+g_0\left(\alpha~e^{-\left(\dfrac{\tau}{\tau_1}\right)^{\beta_1}}+(1-\alpha)~e^{-\left(\dfrac{\tau}{\tau_2}\right)^{\beta_2}}\right)^2    
\label{eq:double_exponential}
\end{equation}
where $\tau_1$ and $\tau_2$ are the two apparent relaxation times with corresponding stretching exponents $\beta_1$ and $\beta_2$, respectively, and $\alpha$ is the relative weight of $\tau_1$ relaxation. E.g. at lower $q$ the cluster scattering dominated and therefore the corresponding relaxation contributed the most to $g_2(q,\tau)$. In the fits shown in Figure \ref{fig:SlowRelax}, $\beta_1$ (fast) and $\beta_2$ (slow) remained at \SI{0.7}{} and \SI{0.3}{}, respectively, while $\alpha$ increased from \SI{0.045}{} at lowest $q$ to \SI{0.49}{} at \SI{0.03}{nm^{-1}}. As before, the mean relaxation times were calculated from the corresponding apparent relaxation time and stretching exponent, $\langle\tau_1\rangle = (\tau_1/\beta_1)\Gamma(1/\beta_1)$ and $\langle\tau_2\rangle = (\tau_2/\beta_2)\Gamma(1/\beta_2)$. All $g_2(q,\tau)$ eventually decayed to \SI{1}{}, indicating insignificant contribution of frozen-in component or dynamical arrest.  

\begin{figure}
    \centering
    \includegraphics[width=0.8\linewidth]{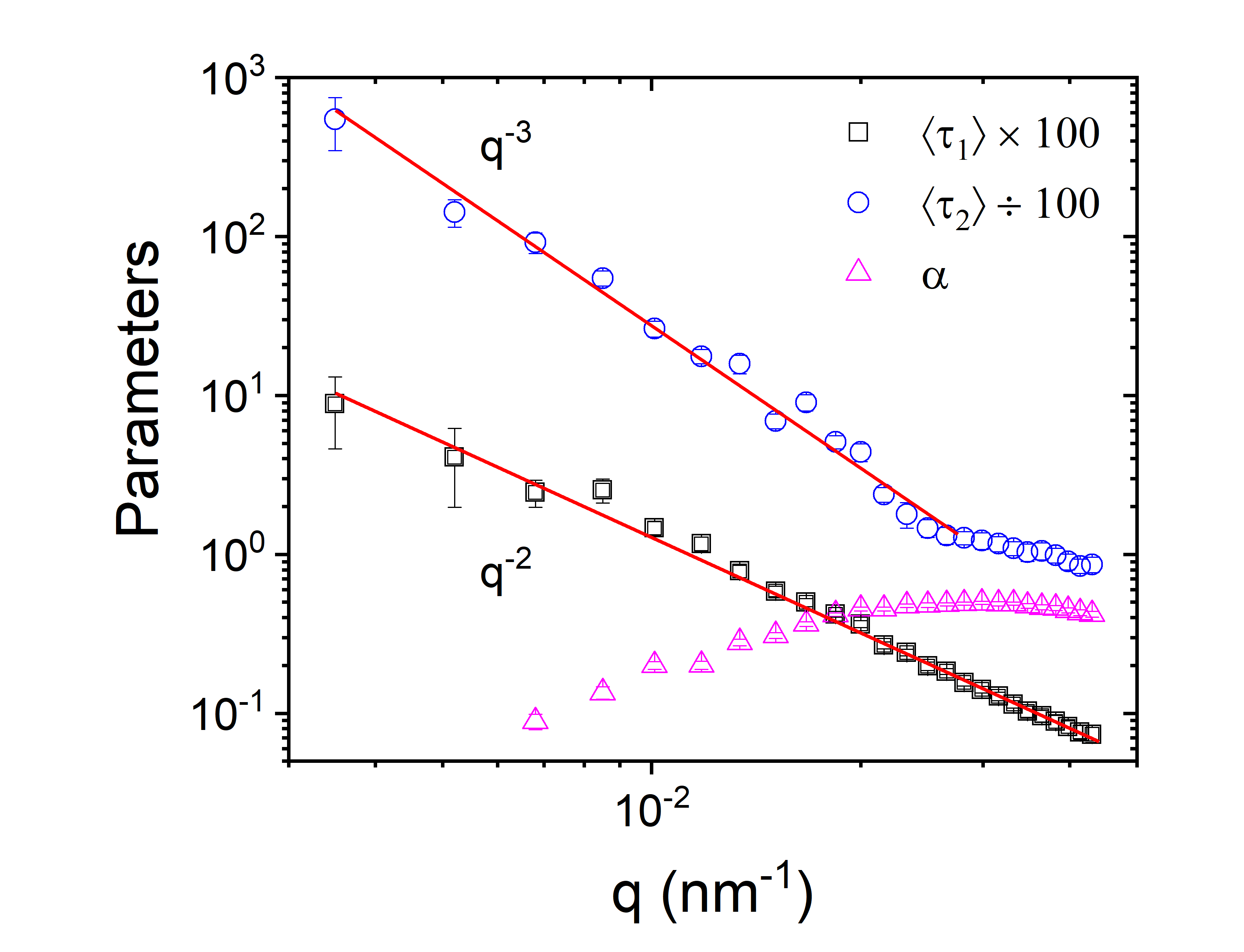}
    \caption{The $q$-dependence of the mean relaxation times, $\langle \tau_1 \rangle$, $\langle \tau_2 \rangle$ and parameter $\alpha$ from the analysis presented in Figure \ref{fig:SlowRelax} (with $\beta_1 \simeq$\SI{0.7}{} and $\beta_2 \simeq$\SI{0.3}{}). Notice the scaling factors associated to  $\langle \tau_1 \rangle$ and $\langle \tau_2 \rangle$, with the latter \SI{5}{-6} orders of magnitude larger. $\langle \tau_2 \rangle$ became nearly $q$ independent below the length scale corresponding to the average interparticle distance. }
    \label{fig:Parameters}
\end{figure}

Figure \ref{fig:Parameters} presents the $q$-dependence of the parameters $\langle\tau_1\rangle$, $\langle\tau_2\rangle$ and $\alpha$ derived from the analysis using Equation (\ref{eq:double_exponential}). The mean fast relaxation time scales as, $\langle\tau_1 \rangle \sim q^{-2}$, which illustrates that individual particles are mobile within the clusters and their bonding is not rigid unlike in irreversible aggregation. However, the corresponding $D \simeq$\SI{0.7}{\mu m^2.s} is much smaller than in the repulsive state together with the stretched exponential relaxation imply constrained diffusion.  The mean slow relaxation time varies as, $\langle\tau_2 \rangle \sim q^{-3}$, until the $q$ range corresponding to the mean interparticle distance and below which it became nearly $q$ independent. This $q$ dependence implies a network-like topology \cite{Cho2020}, consistent with a morphology of percolated clusters that dramatically slowed down the dynamics. The crossover to $q$ independent or $q^0$ behavior has been observed in transient gel networks \cite{Nava2017}, where it was explained by a model involving independent elastically relaxing local units without overall mass transport.       

To cover the \SI{7.5}{} decades in $\tau$ in Figure \ref{fig:SlowRelax} using a linear acquisition would have taken \SI{4e7}{} frames as compared to \SI{2.8e3}{} frames in the bunch-mode. In practice, this may be accomplished by merging three different linear acquisitions of \SI{10000}{} frames with the dead time between frames varied in each set. Even then the linear acquisition would have required a factor \SI{10}{} or more frames and delicate merging of the data sets. In addition, the data volume becomes proportionally larger and computational time increases as its square. While the bunch-mode acquisition is beneficial for covering a broad range of time scales with lower X-ray dose and data volume, it suffers from the sparse sampling of TTCF. This is an issue when investigating slowly relaxing nonergodic systems. This can be noticed in Figure \ref{fig:SlowRelax} for the lowest $q$ values. Therefore, the issue of sufficient ensemble averaging should be addressed when determining the delays between the bunches.

After extensive trials, it turned out that the aperiodic Fibonacci sequence is a reasonable compromise, which is also mathematically tractable and easy to implement via the beamline control software. This sequence arises naturally by searching for the optimal delays between bunches, $t_d$, that should enable good sampling in the $\tau$ landscape. For a bunch of $N$ frames with $\delta t_b$ spacing between frames, each bunch samples $\tau$ from $\delta t_b$ to $N \delta t_b$. If $t_d$ is $N \delta t_b$, the combination of these two bunches samples $\tau$ from  $N \delta t_b$ to $3N \delta t_b$. Then the next bunch should be delayed by $3N \delta t_b$, which will be combined with the first and second bunches, etc. When $t_d$ is exactly a multiple of $N \delta t_b$, $\tau$ at the border will be sampled only by one pair of frames. To address this issue and consequent insufficient ensemble averaging, a multiplicative factor, $\Delta$, was introduced in the setting of optimal bunch delays such that $t_d[n] = (t_d[n-2]+t_d[n-1])~\Delta$. For $\Delta$=\SI{1}{}, the Fibonacci sequence is retrieved, while $\Delta$=\SI{0.5}{} corresponds to equally spaced bunches used in XPCS-Echo measurements. A value of $\Delta <$ \SI{1}{} can be used to achieve better ensemble averaging.    

\section{Conclusion}
The results presented in the previous Section illustrates the effectiveness of bunch-mode XPCS data acquisition particularly for probing systems exhibiting multiple-step broad relaxations, and periodic oscillations. Compared to conventional linear acquisition, bunch-mode aperiodic sequence colossally extends the accessible lag time range while significantly reducing the data volume and the cumulative X-ray dose by roughly a factor \SI{10}{}. The lower accumulated dose aids in minimizing the radiation damage and preserving the sample integrity, especially with biological and soft matter systems. It is particularly well suited for systems exhibiting dynamics over a wide range of time scales like supercooled liquids, glasses and gels \cite{Brown1993,Shen2026,Brackett2026}. However, attention must be paid for ensuring adequate level of ensemble averaging. An aperiodic Fibonacci sequence appears to fulfill this requirement for the systems studied here.  

In the case of periodically driven systems, bunch-mode echo acquisition sequence can be precisely synchronized with a given phase of the external modulation. This approach enables stroboscopic sampling of the dynamics and provides high-resolution measurement of the echoes in the $g_2(q,\tau)$ while discriminating the Doppler shift caused by the periodic motion. The measured echoes in $g_2(q,\tau)$ can be described quantitatively by an analytical expression. The presented method will be useful for probing systems subjected to small amplitude and low frequency oscillatory shear or any other periodic motion \cite{Sala2026}. The protocol of bunch-mode acquisition can be readily extended beyond XPCS to other dynamic scattering methods such as Differential Dynamic Microscopy (DDM) \cite{Edera2021} and multispeckle DLS \cite{Aime2023}.

The flexibility of bunch-mode acquisition may open up new possibilities for designing customized measurement schemes adapted to a broad range of dynamical processes. For instance, adaptive bunch acquisition could be implemented for systems exhibiting aging or non-stationary dynamics, where the temporal evolution slows down continuously. In this case, bunches could be triggered or spaced on the basis of real-time feedback from correlation statistics, pinning the measurements where the dynamic changes are most significant. Another promising direction is the use of phase-tracking acquisition for studying cyclic but anharmonic or multi-frequency driven systems, where the driving force is not perfectly periodic. In this case, a modified echo scheme could lock onto dynamically evolving phases to reconstruct complex time-dependent structural memory.

\section{Data availability}
The original data sets used for the presented analysis are available at https://doi.org/10.15151/esrf-dc-2494118285 . The software tools employed for the data processing can be found at https://gitlab.esrf.fr/id02/id02xpcs . 
\section*{Author Contributions}
\noindent WC, TG and TN designed the project. All authors participated in the research. WC, TN and YC carried out the experiments. WC programmed the correlator to compute $g_2(q,\tau)$ in the bunch-mode acquisition and developed a model to fit the echoes. WC, TN and YC analyzed the data. TN, WC and TG wrote the paper and all authors contributed to the revision. 
\section*{Conflicts of interest}
\noindent There is no conflict of interest to declare.

\section*{Acknowledgments}
\noindent The ESRF is gratefully acknowledged for the provision of synchrotron beam time and financial resources. ID02 beamline staff is thanked for technical support. R. Hino, T. Le Caer, M. Perez, S. Petitdemange and M. Sztucki are acknowledged for MAESTRIO device development and software integration. R. Cerbino and G. Petekidis are thanked for the discussions at the early stage of this development.  
\bibliography{biblio}
\end{document}